\documentclass[11pt]{article}
\usepackage{moriond,epsfig}

\def\Journal#1#2#3#4{{#1} {\bf #2}, #3 (#4)}

\def\PRL{\em Phys. Rev. Lett.}
\def\JETP{{\em JETP}}

\def\be{\begin{equation}}
\def\ee{\end{equation}}
\def\bea{\begin{eqnarray}}
\def\eea{\end{eqnarray}}

\begin{document}
\title{MESOSCOPIC WIGNER CRYSTALLIZATION IN TWO DIMENSIONAL 
LATTICE MODELS}

\author{\underline{JEAN-LOUIS PICHARD}$^1$,
GEORGIOS KATOMERIS$^{1,2}$, FRANCK SELVA$^{1}$}

\address{$^1$Service de Physique de l'Etat Condens\'e, CEA - Saclay, 
91191 Gif sur Yvette cedex, France}
\address{$^2$Department of Physics, University of Ioannina, Greece}

\maketitle

\abstracts{The quantum-classical crossover from the Fermi liquid 
towards the Wigner solid is numerically revisited, considering 
small square lattice models where electrons interact via a 
Coulomb $U/r$ potential. The studies of models without disorder 
and spin and including disorder and spin show that 
the electron solid is formed in two stages, giving rise 
to an intriguing solid-liquid regime at intermediate couplings.}

\section{Lattice model}

 We consider $N$ electrons on $L \times L$ square lattice with 
periodic boundary conditions (BCs). The Hamiltonian reads
\begin{eqnarray}
{\cal H} = \sum_{i,\sigma} (-t \sum_{i'} c^{\dagger}_{i',\sigma}
c_{i,\sigma} + v_i n_{i,\sigma}) +\frac{U}{2} \sum_{i,i'\atop i\neq i'} 
\frac{n_{i,\sigma} n_{i',\sigma'}}
{|i-i'|}+ 2U 
\sum_{i} n_{i,\uparrow}n_{i,\downarrow},
\end{eqnarray}
where $c_{i,\sigma}$ ($c^{\dagger}_{i,\sigma}$) destroys (creates) 
an electron of spin $\sigma$ at the site $i$ and 
$n_{i,\sigma}=c^{\dagger}_{i,\sigma}c_{i,\sigma}$. 
The first terms describe the kinetic energy ($\propto -t$) 
and the random substrate energy (potentials $v_i$ uniformly 
distributed inside $[-W/2,W/2]$). The interaction consists of a 
$U/|i-i'|$ Coulomb repulsion plus a $2U$ Hubbard repulsion. 
$|i-i'|$ is the smallest distance between the sites $i$ 
and $i'$ on a square lattice with periodic BCs. 
In our model, the Coulomb energy to kinetic energy ratio 
$r_s = U/(2t\sqrt{\pi n_e})$ for a filling factor $n_e=N/L^2$. 

${\cal S}$ and ${\cal S}_z$ are the total spin and its component 
along an arbitrary direction $z$. ${\cal H}$ can be written in a 
block-diagonal form, with $N+1$ blocks where $S_z=-N/2,\ldots,N/2$ 
respectively. Assuming $N=4$ and $L=6$, the three blocks with 
$S_z \geq 0$ are diagonalized using Lanczos algorithm to obtain 
the minimum eigenenergy $E_0(S_z)$ of each block. 

\section{Spinless fermions without disorder}

 We begin by studying \cite{katomeris} 
the ground state (GS) of the block ${\cal H} (S_z=2)$ 
when $W=0$.  When $U=0$, the states are $N_H$ plane wave Slater determinants 
(SDs)  $ \prod_{p=1}^4 d^{\dagger}_{k(p)} |0>$, where $d^{\dagger}_{k(p)}$ 
creates a particle in a state of momentum $k(p)= 2\pi (p_x, p_y)/L$ and 
$|0>$ is the vacuum state. The low energy eigenstates are given by: 
(A) $4$ degenerate ground states (GSs)  $|K_0(\beta)>$ ($\beta=1,\ldots,4$) 
of energy $E_0(U=0)=-13 t$ and of momenta  $K_0 \neq 0$; (B) $25$ first 
excitations of energy $E_1 (U=0)=-12 t$. 
(C) $64$ second excitations $|K_2(\alpha)>$ of energy $E_2(U=0)=-11t$; 
(D) $180$ third excitations $|K_3(\alpha)>$ of energy $E_3(U=0)=-10 t$ 
and (E) $384$ fourth excitations of energy $E_4(U=0)=-9t$. We define the 
20 plane wave SDs useful to partly describe the intermediate GS. They are 
given by $4$ plane wave SDs $|K_1(\beta)>$ of energy $-12 t$ where the 
particles have energies $-4t, -3t,-3t,-2t$ respectively, and by $16$ plane 
waves SDs $|K_4(\delta)>$ of energy $-9t$ where the particles have energies 
$-4t,-3t,-2t,0t$ (first set of $8$ SDs) and  $-3t,-3t,-2t,-t$ (second 
set of $8$ other SDs) respectively. Those 20 SDs are directly coupled 
by the pairwise interaction and their momenta are zero 
($K=\sum_{p=1}^4 k(p)=0$). 

When $t=0$, the states are $N_H$ Slater determinants 
$c^{\dagger}_ic^{\dagger}_jc^{\dagger}_kc^{\dagger}_l |0>$ built out 
from the site orbitals. The low energy states are the 
following site SDs: (A) $9$ squares $|S_0(I)>$ ($I=1, \ldots, 9$) of side 
$a=3$ and of energy $E_0(t=0) \approx 1.80 U$; (B) $36$ parallelograms 
$|S_1(I)>$ of sides ($3,\sqrt{10}$) and of energy $\approx 1.85 U$; 
(C) $36$ other parallelograms $|S_2(I)>$ of sides ($\sqrt{10}, \sqrt{10})$ 
and of energy $\approx 1.97 U$ and (D) $144$ deformed squares $|S_3(I)>$ 
obtained by moving a single site of a square $|S_I>$ by one lattice 
spacing and of energy $\approx 2 U$. 
 
%
%

\begin{figure}[ht]  
\vskip.2in
\centerline{\epsfxsize=12cm\epsffile{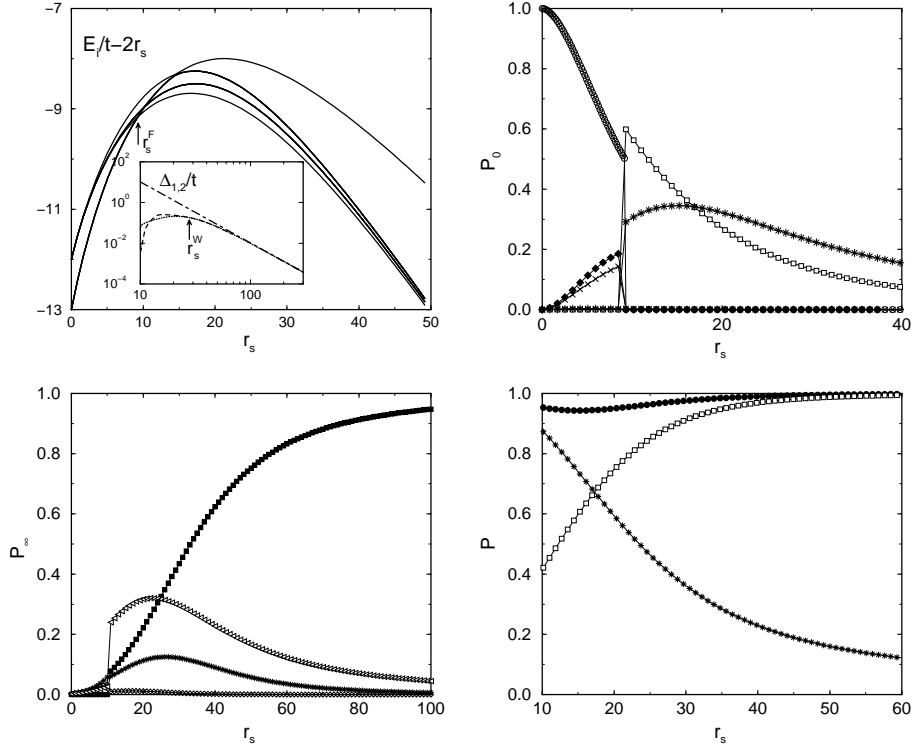}} 
\caption{ UPPER LEFT: Low energy part of the spectrum 
exhibiting a GS level crossing at $r_s^F$. Inset: two first 
level spacings $\Delta_1/t$ (dashed) and $\Delta_2/t$ (dotted) 
which become equal at $r_s^W$ and the perturbative result 
$\Delta_1/t=\Delta_2/t \approx 10392/ r_s^{3}$ valid when 
$r_s \rightarrow \infty$ (dot-dashed). UPPER RIGHT: GS projections 
$P_0(r_s)$ onto a few plane wave SDs, given by the $4 |K_0(\beta)>$ 
(empty circle), the 4 $|K_1(\beta)>$ (empty square), the 64 
$|K_1(\alpha)>$ (filled diamond), the 180 $|K_2(\alpha)>$ ($\times$), 
the 16 $|K_4(\delta)>$ (asterisk) respectively. LOWER LEFT: 
GS projection $P_{\infty}(r_s)$ onto a few site SDs, 
given by the $9$ squares $|S_0(I)>$ (filled square), the $36$ parallelograms 
$|S_1(I)>$ (asterisk), the $36$ parallelograms $|S_2(I)>$ (diamond), 
and the $144$ deformed squares $|S_3(I)>$ (left triangle) respectively.
 LOWER RIGHT: GS projection $P_0^t(r_s)$ (asterisk) 
and $P_{\infty}^t(r_s)$ (empty square) and total GS projection 
$P$ (filled circle) onto the re-orthonormalized basis using the low 
energy eigenvectors of the two limiting bases.} 
\label{CLEAN}
\end{figure}

For the first low energy states, the crossover from the $U=0$ eigenbasis 
towards the $t=0$ eigenbasis is shown in Fig. \ref{CLEAN} (upper left) 
when one increases the ratio $r_s$. If we follow the $4$ GSs $E_0(r_s=0)$ 
($K_0\neq 0$), one can see a first level crossing at $r_s^F \approx 9.3$ 
with a non degenerate state ($K_0=0$) which becomes the GS above $r_s^F$, 
followed by two other crossings with two other sets of $4$ states with 
$K_I \neq 0$. When $r_s$ is large, $9$ states coming from $E_1(r_s=0)$ 
have a smaller energy than the $4$ states coming from $E_0(r_s=0)$. The 
degeneracies ordered by increasing energy become $(1,4,4,4,\ldots)$ 
instead of $(4,25,64,\ldots)$ for $r_s=0$. These $9$ low energy states 
give the $9$ square molecules $|S_0(I)>$ when $r_s \rightarrow \infty$. 
When $r_s^{-1}$ is very small, the first $9$ states correspond to a 
solid molecule free to move on a restricted $3 \times 3$ lattice,
with an effective hopping term $T\propto t r_s^{-3}$. 
This gives $9$ states of  kinetic energy given by $-2T (\cos K_l(I) 
+\cos K_t(I))$ with $K_l(I)=2\pi p_l/3$ and $K_t(I)=2\pi p_t/3$ 
($p_{l,t}=1,2,3$). This structure with degeneracies $1,4,4$ 
respectively and two equal energy spacings $\Delta_1$ and $\Delta_2$ appears 
(inset of Fig. \ref{CLEAN} upper left) when $r_s$ is larger than the 
crystallization threshold $r_s^W \approx 28$. The two characteristic 
thresholds $r_s^F$ (level crossing) and $r_s^W$ (9 first states having the 
structure of the spectrum of a single solid molecule free to move 
on a $3 \times 3$ square lattice) can also be detected by other methods 
given in Ref. [1].  

To understand further the nature of the intermediate GS between 
$r_s^F$ and $r_s^W$, we have projected the GS wave functions 
$|\Psi_0(r_s)>$ over the low energy eigenvectors of the two eigenbases 
valid for $U/t=0$ and for $t/U=0$ respectively. Let us begin with 
the $U=0$ eigenbasis. Below $r_s^F$, each of the 4 GSs 
$|\Psi_0^{\alpha}(r_s)>$ with $K_0\neq 0$ has still a large projection  
$P_0(r_s,0)=\sum_{\beta=1}^4 |<\Psi_0^{\alpha}(r_s)| K_0({\beta})>|^2 $
over the $4$ non interacting GSs. There is no projection over the $25$ first 
excitations and smaller projections $P_0(r_s,2)$ and $P_0(r_s,3)$ over the 
$64$ second and $180$ third excitations of the non interacting system. 
Above $r_s^F$, the non degenerate GS with $K_0=0$ has a large projection 
\begin{equation}
P_0(r_s,1)=\sum_{\beta=1}^4 |<\Psi_0(r_s)| K_1({\beta})>|^2 
\end{equation}
which is equally distributed over the $4$ excitations $|K_1(\beta)>$ 
of momentum $K_1=0$ and a second significant contribution 
\begin{equation}
P_0(r_s,4)=\sum_{\delta=1}^{16} |<\Psi_0(r_s)| K_4({\delta})>|^2 
\end{equation}
given by its projection onto the $16$ mentionned plane wave SDs 
$|K_4({\delta})>$ of energy $-9t$. Above $r_s^F$, its projections 
onto the $4$ $|K_0({\beta})>$, the $21$ other first excitations 
and the second and third excitations of the non interacting system 
are zero or extremely negligible. The total GS projection 
$P_0^t(r_s)=P_0(r_s,1)+P_0(r_s,4)$ onto the $4$ $|K_1({\beta})>$ and 
$16$ $|K_4({\delta})>$ is given in Fig. \ref{CLEAN} (lower right) 
when $r_s > r_s^F$. This shows us that a large part of the system 
remains an excited liquid above $r_s^F$. 

 We now study the GS projections $P_{\infty}$ onto the $t=0$ eigenbasis.  
The GS projection 
\begin{equation}
P_{\infty}(r_s,0)=\sum_{I=1}^9 |<\Psi_0^{\alpha}(r_s)| S_0(I)>|^2 
\end{equation}
onto the $9$ square site SDs $|S_0(I)>$ is given in Fig. \ref{CLEAN} (lower 
left), together with the GS projection $P_{\infty}(r_s,J)$ onto the 
site SDs corresponding to the $J^{th}$ degenerate low energy excitations 
of the $t=0$ system.  The total GS projection 
$P_{\infty}^t(r_s)= \sum_{p=0}^3 P_{\infty}(r_s,p) $ 
onto the $9$ squares $|S_0(I)>$, the $36$ parallelograms $|S_1(I)>$, 
the $36$ other parallelograms $|S_2(I)>$ and the $144$ deformed squares 
$|S_3(I)>$ is given in Fig. \ref{CLEAN} (lower right) when $r_s > r_s^F$. 
This shows us that the ground state begins to be a floppy solid also 
above $r_s^F$. 

The site SDs and plane wave SDs are not orthonormal. After 
re-orthonormalization, the total projection $P$ of $|\Psi_0(r_s)>$ 
over the subspace spanned by the $4$ $|K_1(\beta)>$ and $16$ $|K_4(\delta)>$ 
and $225$ site SDs of lower electrostatic energies are given in 
Fig. \ref{CLEAN} (lower right), showing that $|\Psi_0(r_s)>$ is almost 
entirely located inside this very small part of a huge Hilbert space 
for intermediate $r_s$, spanned by low energy SDs of different nature, and 
adapted to describe a solid entangled with an  excited liquid. 

\section{Magnetization and polarization energies in presence 
of a random substrate}
\label{sec:mag}

We now consider weakly disordered samples when the spin degrees of 
freedom are included \cite{selva}. Their role and the consequences 
of an applied parallel magnetic field which aligns only the spins 
without inducing orbital effects, have been the subject of Ref. [2]. 
The study of a statistical ensemble of samples with $W=5$, $N=4$ and $L=6$ 
provides complementary signatures of a particular intermediate behavior. 
Let us note $M$ the fraction of clusters with $S=1$ at $B=0$, 
$Q_2=E_0(S_z=2)-E_0(S_z=0)$ and $Q_1 =E_0(S_z=1)-E_0(S_z=0)$ the 
Zeeman energies necessary to yield $S=2$ and $S=1$ respectively for a 
cluster with $S=0$. 

\begin{figure}
\centerline
{
\epsfxsize=8cm
\epsffile{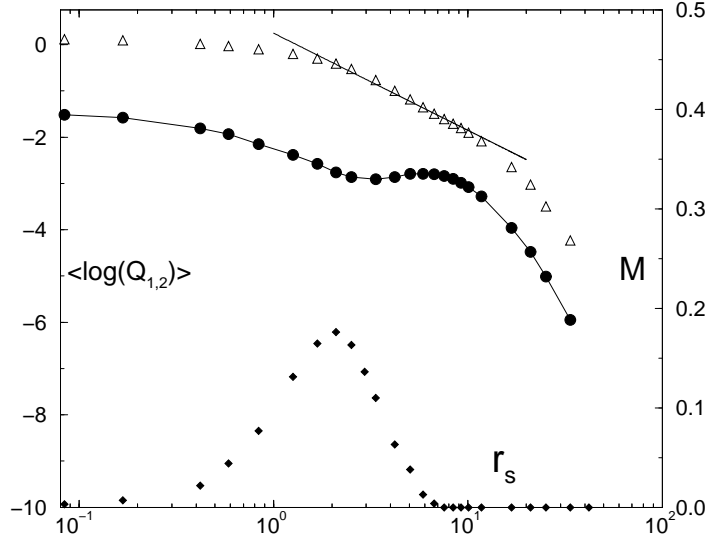}
}
\vspace{0.5cm}
\caption
{
As a function of $r_s$, fraction $M$ of clusters with $S=1$ 
at $B=0$ (filled diamond, right scale),  partial $<\log Q_1>$ 
(filled circle, left scale) and total $<\log Q_2>$ (empty triangle, 
left scale) energies required to polarize $S=0$ clusters to $S=1$ and $S=2$ 
respectively. The straight line corresponds to $0.25 - 2 \log r_s$. 
}
\label{magnetization-fig2}
\end{figure}

 In Fig. \ref{magnetization-fig2}, $M$ is given as a function of $r_s$. 
One can see a first threshold at $r_s \approx 0.35$ where the interaction 
can drive $S=1$ in certain samples. Above a second threshold $r_s^{FS}$, 
$M$ regularly decreases to reach a zero value at a third threshold 
$r_s^{WS} \approx 9$ where an antiferromagnetic square molecule is formed. 
The ensemble averages $<\log Q_1>$ and $<\log Q_2>$ (without taking into 
account the $S=1$ spontaneously magnetized clusters) define the typical 
fields $B$ necessary to yield $S=1$ or $S=2$ in a $S=0$ cluster. 
In Fig. \ref{magnetization-fig2}, one can see an intermediate regime 
again for $ r_s^{FS} < r_s < r_s^{WS}$ where $<\log Q_1>$ becomes roughly 
independent of $r_s$, while $Q_2 \propto r_s^{-2}$.

\section{Conclusion}
\label{sec:conclusion}

 One concludes that mesoscopic Wigner crystallization proceeds in two 
stages. In a clean system, a minimal description of the intermediate GS 
requires to combine the low energy states of the two limiting eigenbases. 
In this sense, the intermediate GS is neither solid, nor liquid, but 
rather the quantum superposition of those two states of matter. This is 
strongly reminiscent of the conjecture proposed by Andreev and Lifshitz 
\cite{AL} for the quantum melting of a solid. 
A path integral Monte Carlo approach \cite{filinov} has recently shown 
that a few electrons 
confined in a  harmonic trap crystallize also in two stages, firstly via a 
radial ordering of electrons on shells and secondly via the freezing of the 
intershell rotation. As one can see, a two stage crystallization is not 
only characteristic of a mesoscopic harmonic trap, but also occurs in a 
mesoscopic $2d$ torus. To add a random substrate defavors the liquid state.  
The magnetization gives also the signature of an intermediate regime 
where Stoner ferromagnetism is defavored by Wigner antiferromagnetism.


\section*{References}
\begin{thebibliography}{99}

\bibitem{katomeris} G. Katomeris and J.-L. Pichard, cond-mat/0012213, 
submitted to {\em Phys. Rev. Lett.}.

\bibitem{selva} F. Selva and J.-L. Pichard, cond-mat/0012015, 
submitted to {\em Europhys. Lett.}.

\bibitem{AL} A. F. Andreev and I. M. Lifshitz, \Journal{\JETP}{29}{1107}{1969}.

\bibitem{filinov} A. V. Filinov, M. Bonitz and Yu. E. Lozovik, 
\Journal{\PRL}{86}{3851}{2001}.

\end{thebibliography}
\end{document}